\documentclass[english]{ppgeesa}

\usepackage[utf8]{inputenc}
\usepackage{graphicx}
\usepackage{amsmath}
\usepackage{amssymb}
\usepackage{algorithm}
\usepackage{algpseudocode}
\usepackage{url}
\usepackage{hyperref}
\hypersetup{
    colorlinks=true,
    linkcolor=blue,
    citecolor=blue,
    urlcolor=blue
}
\begin{document}

\title{CAMP: Cumulative Agentic Masking and Pruning for
Cross-Turn PII Protection in Multi-Turn LLM Conversations}

\author{Aman Panjwani\\
{\small Independent Researcher}\\
{\small amanpanjavani111@gmail.com}}

\maketitle
\thispagestyle{empty}\pagestyle{empty}

\makeatletter
\renewenvironment{abstract}{%
  \par\vspace{1ex}%
  {\centering\normalfont\textbf{Abstract}\par}%
  \vspace{0.5ex}%
  \small\noindent\ignorespaces
}{%
  \par\vspace{1ex}%
}
\makeatother

\begin{abstract}
The deployment of Large Language Models (LLMs) in agentic,
multi-turn conversational settings has introduced a class of
privacy vulnerabilities that existing protection mechanisms are
not designed to address. Current approaches to Personally
Identifiable Information (PII) masking operate on a per-turn
basis, scanning each user message in isolation before forwarding
sanitized text to the model. While effective against direct
identifier leakage within a single message, these methods are
fundamentally stateless and fail to account for the compounding
privacy risk that emerges when PII fragments accumulate across
conversation turns. A user who separately discloses their name,
employer, location, and medical condition across several messages
has revealed a fully re-identifiable profile, yet no individual
message would trigger a per-turn masker. We formalize this
phenomenon as Cumulative PII Exposure (CPE) and propose CAMP
(Cumulative Agentic Masking and Pruning), a cross-turn privacy
protection framework that maintains a session-level PII registry,
constructs a co-occurrence graph to model combination risk between
entity types, computes a CPE score after each turn, and triggers
retroactive pseudonymization of conversation history when the
score crosses a configurable threshold. We evaluate CAMP on four
synthetic multi-turn scenarios spanning healthcare, hiring,
finance, and general conversation, demonstrating that per-turn
baselines expose re-identifiable profiles that CAMP successfully
neutralizes while preserving full conversational utility.
\end{abstract}


\section{Introduction}

The rapid proliferation of Large Language Models (LLMs) in
production environments has given rise to a new class of
conversational systems known as agentic assistants,
applications that maintain context across multiple interaction
turns to complete complex, goal-oriented tasks on behalf of
users~\cite{anthropic2024agents}. From healthcare triage
assistants and HR onboarding bots to financial advisors and
customer support agents, these systems routinely process
sensitive user input as part of their normal operation. Unlike
single-turn query systems, agentic frameworks such as
LangChain~\cite{langchain2023} explicitly preserve and forward
the full conversation history to the model on every turn,
allowing the LLM to reason over the complete context of a
session. This architectural property, essential for coherent
multi-turn reasoning, simultaneously creates an expanding
surface for privacy risk.

Existing approaches to protecting Personally Identifiable
Information (PII) in LLM pipelines focus almost exclusively
on the individual message. Tools such as Microsoft
Presidio~\cite{presidio2021} and hybrid systems such as
PAPILLON~\cite{siyan2024papillon} intercept each user message
before it reaches the model, apply Named Entity Recognition
(NER) or pattern-based detection to identify sensitive spans,
and replace them with typed placeholders such as
\texttt{<PERSON>} or \texttt{<EMAIL>}. These systems operate
without any memory of prior turns. Each message is evaluated
independently, and the masking decision is made solely on the
content of that message in isolation.

This design assumption, that privacy risk is contained within
individual messages, does not hold in multi-turn conversations.
Privacy leakage in agentic settings follows a fundamentally
different pattern: it is gradual, distributed, and
combinatorial. Consider a user interacting with a banking
assistant over eight turns. In turn~1 they ask a general
question. In turn~2 they provide their name. In turn~4 they
mention their city. In turn~5 they reveal their employer.
In turn~7 they disclose their annual income. No single message
in this exchange would trigger a conventional per-turn masker,
yet by turn~7 the conversation history forwarded to the LLM
contains a profile that is substantially re-identifiable:
name, location, employer, and income together constitute a
combination that uniquely characterises the individual. This
is not an edge case. It is the normal pattern of how humans
disclose information across a conversation.

This phenomenon, which we term \textbf{Cumulative PII
Exposure (CPE)}, represents an important vulnerability in
deployed agentic LLM systems that has not yet been formally
studied. The risk does not originate from any single
disclosure but from the intersection of multiple
quasi-identifiers that, when combined, cross the threshold
of re-identification. This is closely related to the
classical mosaic effect in information security, wherein
individually innocuous pieces of information compose a
sensitive profile when aggregated~\cite{nissenbaum2004}.
In the context of multi-turn LLM conversations this effect
is amplified by the fact that the conversation history is
explicitly maintained and forwarded to an external model,
meaning the accumulated profile represents a concrete data
exposure at every subsequent turn.

We propose \textbf{CAMP (Cumulative Agentic Masking and
Pruning)}, a cross-turn privacy protection framework designed
to address CPE in multi-turn LLM conversations. CAMP
introduces five coordinated components that operate as a
middleware layer between the user and the LLM. First, a
per-turn extractor identifies PII entities in each message
using pattern-based recognition. Second, a session-level
registry maintains a stateful record of all entities detected
across the conversation, a component that has no equivalent
in any existing per-turn system. Third, a co-occurrence graph
models the combination risk between entity types, capturing
the intuition that a location entity appearing in the same
session as a salary entity carries significantly higher
privacy risk than either entity in isolation. Fourth, a
retroactive pseudonymizer intervenes on the conversation
history when the CPE score crosses a configurable threshold,
replacing real PII fragments with consistent synthetic
identities before they are forwarded to the LLM. Fifth, a
de-masking layer intercepts the LLM response and restores
the original identities before the response is returned to
the user, ensuring that the end user experience remains
fully coherent while the external model never processes real
PII. The pseudonym map is maintained locally and never
transmitted to the external model.

To evaluate CAMP we construct four synthetic multi-turn
scenarios covering the domains of healthcare, hiring,
finance, and general conversation, designed to reflect
realistic disclosure patterns where no individual turn
triggers per-turn masking but the accumulated context
creates measurable re-identification risk. We demonstrate
that standard per-turn baselines expose the full accumulated
profile to the LLM across all four scenarios, while CAMP
detects the combination risk and intervenes before the
re-identification threshold is crossed, preserving full
conversational utility throughout.

The contributions of this paper are as follows:
\begin{itemize}
\item We formally define Cumulative PII Exposure (CPE) as a
session-level privacy risk metric for multi-turn LLM
conversations, the first such formalization to our knowledge.
\item We introduce a co-occurrence graph model that captures
combination risk between PII entity types, enabling detection
of re-identification risk that emerges from entity co-presence
rather than individual entity sensitivity.
\item We present CAMP, a retroactive pseudonymization
framework that rewrites conversation history when cumulative
risk crosses a threshold, a capability not present in
existing per-turn masking systems.
\item We design a local de-masking layer that restores
original identities in LLM responses before returning them
to the user, completing a full privacy-preserving round trip
without degrading conversational utility.
\item We evaluate CAMP on four synthetic multi-turn scenarios
and demonstrate its effectiveness at neutralizing
combination-driven PII exposure while preserving
conversational utility.
\end{itemize}

\section{Background and Related Work}

The problem of protecting personally identifiable information
in text-based AI systems has attracted growing attention
across NLP, security, and privacy research communities. We
review three bodies of work that together provide the
foundation for our approach: PII detection and masking in
NLP pipelines, privacy risks specific to LLM inference, and
emerging work on privacy in agentic and multi-turn systems.

\subsection{PII Detection and Masking}

The problem of detecting and removing personally identifiable
information from text has a long history in NLP, originating
primarily in the medical and legal domains where
de-identification of clinical notes and court documents was
required for research use~\cite{lison2021anonymisation}.
Early approaches relied on rule-based pattern matching and
regular expressions to identify structured identifiers such
as social security numbers, phone numbers, and dates. These
methods remain effective for well-formatted structured PII
but face limitations on contextually embedded or implicit
identifiers.

The introduction of Named Entity Recognition (NER) models
brought significant improvements in detecting unstructured
PII such as person names, organizations, and locations.
Modern PII masking pipelines typically combine NER-based
detection with pattern matching. Microsoft
Presidio~\cite{presidio2021} is the most widely adopted
open-source implementation of this approach, supporting over
twenty entity types and providing a reversible anonymization
layer through synthetic replacement. Presidio operates at
the message level and carries no state between invocations.
Early work by Lukas et al.~\cite{lukas2023analyzing}
demonstrated that language models can leak PII from training
data under targeted extraction attacks, establishing the
empirical basis for the threat that per-turn maskers aim
to address.

More recent work has explored LLM-based approaches to PII
detection and masking. PAPILLON~\cite{siyan2024papillon}
proposes a hybrid architecture in which a local LLM filters
and transforms PII before forwarding partially sanitized
queries to a remote API-based model, reducing privacy
leakage from 100\% to 7.5\% while maintaining response
quality. The Adaptive PII Mitigation
Framework~\cite{adaptivepii2025} introduces
regulation-aware masking that adjusts anonymization
strategies based on jurisdiction-specific requirements such
as GDPR and CCPA, achieving strong detection performance
across a range of entity types. INTACT~\cite{intact2024}
replaces sensitive spans with semantically general but
truth-preserving alternatives using instruction-tuned
open-weights models, reducing re-identification risk with
only marginal utility loss. Each of these contributions
advances the state of the art in per-message privacy
protection, and our work builds on their foundations by
extending the scope of analysis to the session level.

Table~\ref{tab:comparison} summarises the key distinctions
between existing systems and CAMP.

\begin{table}[htbp]
\begin{center}
\caption{Comparison of Existing PII Masking Approaches}
\label{tab:comparison}
\renewcommand{\arraystretch}{1.6}
\begin{tabular}{p{1.5cm} p{1.5cm} p{1.2cm} p{2.5cm}}
\hline
\textbf{System} & \textbf{Scope} & \textbf{Stateful}
& \textbf{Strategy} \\
\hline
Presidio  & Per-turn & No  & NER + regex      \\[4pt]
PAPILLON  & Per-turn & No  & LLM filter       \\[4pt]
INTACT    & Per-turn & No  & Generalization   \\[4pt]
Adaptive  & Per-turn & No  & Policy-driven    \\[4pt]
CAMP      & Session  & Yes & Pseudonymization \\
\hline
\end{tabular}
\end{center}
\end{table}

\subsection{Privacy Risks in LLM Inference}

A parallel line of research has examined the privacy risks
that arise during LLM inference. Carlini et
al.~\cite{carlini2021extracting} first demonstrated verbatim
extraction of training data from production LLMs, and Nasr
et al.~\cite{nasr2023scalable} showed this scales to modern
model sizes with a substantial fraction of extracted content
being authentic. Staab et al.~\cite{staab2024beyond}
demonstrated that LLMs can infer sensitive personal
attributes including location, age, and profession from
text that contains no explicit PII, using only contextual
reasoning. This observation motivates our focus on
combination risk, as it establishes that even indirect
disclosure accumulation can lead to attribute inference by
a capable model.

PII-Scope~\cite{nakka2024piiscope} provides a comprehensive
benchmark for PII extraction attacks, demonstrating that
with sophisticated adversarial strategies PII extraction
rates can increase substantially compared to single-query
baselines. The adversarial anonymization framework of Staab
et al.~\cite{staab2024adversarial} demonstrated that
iterative LLM-based rewriting can reduce re-identification
risk beyond what rule-based tools achieve, and
SEAL~\cite{seal2025} extended this with self-refining
anonymizers via adversarial distillation. Recent surveys
of LLM privacy~\cite{surveypii2025} and
utility-preserving anonymization
frameworks~\cite{rupta2025} have confirmed that the
privacy-utility tradeoff is a central challenge in text
anonymization research. These findings collectively
establish that LLMs actively process and retain sensitive
information, making the content of every prompt a
privacy-relevant event and reinforcing the importance of
controlling what reaches the model in the first place.

\subsection{Privacy in Agentic and Multi-Turn Systems}

The privacy implications of agentic LLM systems have
received comparatively limited attention relative to the
broader LLM privacy literature.
AgentLeak~\cite{agentleak2026} provides a valuable recent
contribution by introducing the first comprehensive
benchmark for privacy leakage in multi-agent LLM systems,
covering scenarios across healthcare, finance, legal, and
corporate domains. AgentLeak identifies coordination
attacks as a distinct threat in multi-agent architectures,
where sensitive information can propagate through
inter-agent delegation in ways that single-agent privacy
controls do not anticipate. The Privacy Guard
framework~\cite{privacyguard2025} acknowledges that long
conversational histories can make sensitive data inferrable
even when no single message contains explicit PII, but does
not provide a formal metric or retroactive intervention
mechanism.

Contextual Integrity theory~\cite{nissenbaum2004},
originally developed in the context of information ethics,
offers a principled framework for reasoning about
appropriate information flows. Mireshghallah et
al.~\cite{mireshghallah2023quantifying} introduced
contextual integrity benchmarks for LLMs, showing that
models frequently violate contextual norms in their
outputs. Our CPE formulation is philosophically aligned
with this perspective: we treat the appropriateness of
forwarding information to an external model as dependent
not on the sensitivity of any individual entity in
isolation but on the combination of entities that have
accumulated across the session.

Taken together, the literature on per-turn masking,
inference-time privacy risks, and agentic privacy points
toward a setting that has not yet been directly studied:
the progressive, cross-turn accumulation of PII within
a single conversational session and its implications for
what is forwarded to an external LLM. CAMP is our
contribution toward addressing this setting.

\section{Cumulative PII Exposure: Definition and
Formalization}

In this section we introduce the formal foundations of
CAMP. We define the multi-turn conversation model, the
PII entity model, the co-occurrence graph, and the CPE
score that drives the masking decision. We conclude with
the threat model under which CAMP operates.

\subsection{Conversation Model}

We model a multi-turn LLM conversation as an ordered
sequence of user messages $\{m_0, m_1, \ldots, m_T\}$
where $m_t$ denotes the message at turn $t$. In agentic
frameworks such as LangChain~\cite{langchain2023}, the
context window forwarded to the LLM at turn $t$ includes
the full conversation history up to that point. This means
any unmasked PII from earlier turns is re-exposed to the
external model on every subsequent call, creating an
expanding privacy surface as the session grows.

\subsection{PII Entity Model}

The entity types considered by the system are:
\textsc{Person}, \textsc{Location}, \textsc{Organization},
\textsc{Email}, \textsc{Phone}, \textsc{Date of Birth},
\textsc{SSN}, \textsc{Medical Condition}, and
\textsc{Salary}. Each entity type $e$ is assigned a base
sensitivity weight $w(e) \in (0, 1]$ reflecting its
standalone re-identification risk. Direct identifiers
carry higher weight than quasi-identifiers as shown in
Table~\ref{tab:weights}.

\begin{table}[htbp]
\begin{center}
\caption{Entity Type Sensitivity Weights}
\label{tab:weights}
\renewcommand{\arraystretch}{1.3}
\begin{tabular}{p{2.8cm} p{1.0cm} p{2.8cm} p{1.0cm}}
\hline
\textbf{Entity Type} & \textbf{w(e)} &
\textbf{Entity Type} & \textbf{w(e)} \\
\hline
SSN               & 1.00 & Person         & 0.60 \\
Credit Card       & 1.00 & Salary         & 0.60 \\
Date of Birth     & 0.90 & Location       & 0.50 \\
Medical Condition & 0.85 & IP Address     & 0.50 \\
Email Address     & 0.80 & Organization   & 0.30 \\
Phone Number      & 0.75 &                &      \\
\hline
\end{tabular}
\end{center}
\end{table}

The accumulated entity set after turn $t$ is the union of
all entity types detected across turns 0 through $t$.

\subsection{The Co-occurrence Graph}

Privacy risk in a multi-turn session cannot be evaluated
by summing individual entity weights alone. A location
entity appearing in isolation is a weak quasi-identifier.
The same location entity appearing alongside a salary
disclosure and an employer name creates a combination that
is substantially more dangerous. We model this combination
risk using a co-occurrence graph where each node represents
a PII entity type seen in the session and an edge connects
two nodes if both entity types have appeared within the
same session. The graph grows monotonically as the
conversation progresses, with new nodes and edges added
as new entity types are detected across turns.

Each node is assigned a combination amplifier:

\begin{equation}
f(v) = 1 + \alpha \cdot \deg(v)
\end{equation}

where $\deg(v)$ is the number of edges incident to that
node and $\alpha \in (0, 1)$ controls amplifier
sensitivity. A node with no edges contributes only its
base weight. A node connected to several other entity
types receives a higher amplifier, reflecting its elevated
role in combination risk.

\subsection{CPE Score}

The Cumulative PII Exposure score at turn $t$ is defined
as the sum of amplified entity weights across all
accumulated entity types:

\begin{equation}
\text{CPE}(t) = \sum_{v} w(v) \cdot \left(1 + \alpha
\cdot \deg(v)\right)
\end{equation}

The score is monotonically non-decreasing across turns
since entity types are never removed from the session
registry. It grows superlinearly when new entity types
form edges with existing ones, as both the new node and
all its neighbors receive increased amplification. This
superlinear growth formally captures the intuition that
combination risk escalates faster than individual
disclosures.

\subsection{Pseudonymization and De-masking}

When $\text{CPE}(t) \geq \tau$, CAMP triggers retroactive
pseudonymization of the full conversation history. Each
real PII value in the session registry is replaced with
a consistent synthetic substitute generated locally using
the Faker library. The mapping between real and synthetic
values is stored in a local pseudonym table that is never
transmitted to the external model. The rewritten history
is forwarded to the LLM in place of the original. When
the LLM returns a response, a de-masking layer applies
the reverse mapping to restore original identities before
the response is shown to the user, completing a full
privacy-preserving round trip.

\subsection{Threat Model}

We consider a setting in which a user interacts with an
agentic LLM assistant that forwards conversation history
to a third-party hosted LLM API. The adversary is any
entity with access to the API request logs who may attempt
re-identification by combining PII fragments present
across the conversation history. The user is assumed to
disclose PII naturally as part of legitimate task
completion. The operator deploys CAMP as a trusted
middleware layer and the LLM is treated as an untrusted
external service. The re-identification threshold $\tau$
is a configurable parameter evaluated across
$\tau \in \{1.5, 2.0, 2.5\}$ in our experiments.

\section{The CAMP Framework}

CAMP operates as a middleware layer positioned between
the user and the external LLM. It intercepts every user
message before it reaches the model, maintains a stateful
record of the session, and intervenes on the conversation
history when the accumulated privacy risk crosses a
configurable threshold. The framework consists of five
components described in the following subsections.

\subsection{Per-Turn PII Extractor}

The extractor processes each incoming user message and
identifies PII entities using a combination of Named
Entity Recognition and pattern-based rules. In our
implementation we use Microsoft
Presidio~\cite{presidio2021} with custom domain-specific
recognizers that extend the base detection capability
with patterns for salary mentions, account number
fragments, and street addresses. The extractor returns
a list of detected entity types and their corresponding
text spans for the current turn. This component operates
entirely locally and no user message is forwarded to any
external service at this stage.

\subsection{Session Registry}

The session registry maintains a stateful record of every
PII entity detected across all turns in the conversation.
It stores the entity type, the original text value, and
the turn index at which each entity was first observed.
The registry serves as the single source of truth for all
downstream components. It accumulates monotonically across
the session and is never transmitted to the external
model. This is the component that fundamentally
distinguishes CAMP from per-turn masking systems, which
discard all entity information after each turn.

\subsection{Co-occurrence Graph and CPE Scorer}

\begin{figure}[htbp]
\centering
\includegraphics[width=\columnwidth]{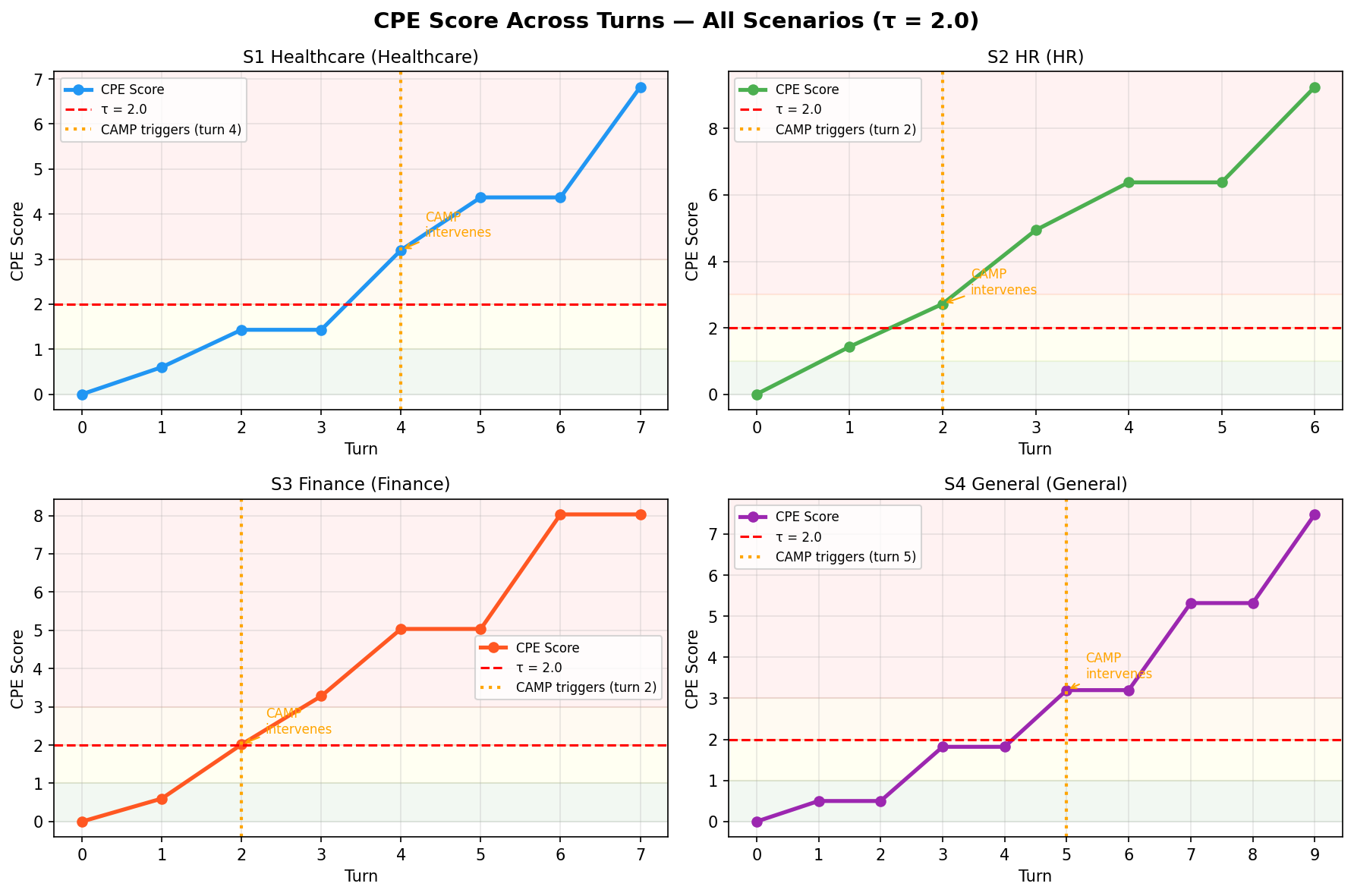}
\caption{CPE score across turns for all four scenarios
at $\tau = 2.0$. The dashed red line marks the threshold.
Orange dotted lines mark CAMP intervention points.}
\label{fig:cpe_curves}
\end{figure}

After each turn the co-occurrence graph is updated with
any new entity types detected by the extractor. New nodes
are added for entity types appearing for the first time
and new edges are formed between all pairs of entity types
that now co-exist in the session registry. The CPE score
is then recomputed using equation~(2). If the updated
score remains below the threshold $\tau$ the current
message is forwarded to the LLM without modification.
If the score meets or exceeds $\tau$ the pseudonymizer
is triggered before any message is sent.
Figure~\ref{fig:cpe_curves} illustrates the CPE score
rising across turns across all four evaluation scenarios.

\subsection{Retroactive Pseudonymizer}

When the CPE threshold is crossed the pseudonymizer
rewrites the entire conversation history stored in the
session registry. For each real PII value it generates
a consistent synthetic substitute using the Faker
library. Person names are replaced with plausible
alternative names, locations with plausible alternative
cities, organizations with plausible alternative company
names, and numerical values such as salary with
randomized alternatives within a realistic range. The
same real value always maps to the same synthetic
substitute within a session, ensuring referential
consistency across turns. The rewritten history is
assembled as a new context window and forwarded to the
LLM in place of the original.

\subsection{De-masking Layer}

When the LLM returns a response the de-masking layer
scans the response text for occurrences of synthetic
substitutes and replaces them with the corresponding
original values using the reverse of the pseudonym map.
The user therefore receives a response that refers to
their real name, real institution, and real location,
while the external model processed only synthetic
identities throughout. The pseudonym map is held in
memory for the duration of the session and is discarded
when the session ends.

\subsection{Hard Block Rules}

Independent of the CPE score, certain entity types are
never forwarded to the external model under any
circumstances. These include SSN, credit card numbers,
bank account numbers, and IBAN codes. When any of these
entity types is detected by the extractor the
corresponding text span is replaced with a typed
placeholder such as \texttt{[BLOCKED]} before the
message is sent. This rule applies from turn~0 and
cannot be overridden by threshold configuration. The
distinction between hard-blocked and pseudonymized
entities reflects the principle that some identifiers
carry irreversible re-identification risk that no amount
of session-level context justifies transmitting
externally.

\subsection{Processing Pipeline}

Algorithm~\ref{alg:camp} summarizes the per-turn
processing pipeline of CAMP.

\begin{algorithm}
\caption{CAMP Per-Turn Processing}
\label{alg:camp}
\begin{algorithmic}[1]
\Require User message $m_t$, registry $\mathcal{R}$,
         graph $\mathcal{G}$, threshold $\tau$
\Ensure  Response $r$ returned to user
\State Extract PII entities from $m_t$ locally
\State Apply hard block rules to $m_t$
\State Update registry $\mathcal{R}$ with new entities
\State Update co-occurrence graph $\mathcal{G}$
\State Compute $\text{CPE}(t)$ using Equation~(2)
\If{$\text{CPE}(t) \geq \tau$}
    \State Generate pseudonym map $\mathcal{P}$
    \State Rewrite full history $\mathcal{W}_t$
           using $\mathcal{P}$
\EndIf
\State Forward context window to LLM
\State Receive response $r_{\text{pseudo}}$
\State Apply reverse map $\mathcal{P}^{-1}$
       to $r_{\text{pseudo}}$
\State \Return de-masked response $r$ to user
\end{algorithmic}
\end{algorithm}

\section{Experiments}
\label{sec:experiments}

We evaluate CAMP on four synthetic multi-turn
conversation scenarios designed to reflect realistic
patterns of gradual PII disclosure across domains. We
compare CAMP against a per-turn Presidio baseline across
three metrics: entities exposed to the external model,
CPE score at session end, and the turn at which
intervention occurs.

\subsection{Experimental Setup}

All experiments are conducted on synthetic conversations
constructed to represent natural human disclosure patterns
where no individual turn contains sufficient PII to
trigger a per-turn masker, yet the accumulated session
context creates measurable re-identification risk. Each
scenario is run under three threshold settings
$\tau \in \{1.5, 2.0, 2.5\}$ to characterize sensitivity
to threshold selection. The combination amplifier
parameter is fixed at $\alpha = 0.3$ across all
experiments. The external LLM used in all scenarios is
Claude Sonnet~4.6 deployed via Azure AI Foundry.

\subsection{Scenarios}

Table~\ref{tab:scenarios} summarizes the four evaluation
scenarios. Each scenario is characterized by its domain,
number of turns, the turn at which CAMP intervenes under
$\tau = 2.0$, and the entity types that accumulate
across the session.

\begin{table}[htbp]
\begin{center}
\caption{Evaluation Scenarios}
\label{tab:scenarios}
\renewcommand{\arraystretch}{1.3}
\begin{tabular}{p{0.8cm} p{1.2cm} p{0.6cm} p{1.2cm}
p{2.6cm}}
\hline
\textbf{ID} & \textbf{Domain} & \textbf{Turns} &
\textbf{Trigger} & \textbf{Entity Types} \\
\hline
S1 & Healthcare & 8  & Turn 4 &
Person, Location, DOB, Medical, Email \\
S2 & HR/Hiring  & 7  & Turn 2 &
Person, Location, Salary, Age, Email \\
S3 & Finance    & 8  & Turn 2 &
Person, Location, Salary, Account, SSN \\
S4 & General    & 10 & Turn 5 &
Person, Location, Org, Medical, Email, Salary \\
\hline
\end{tabular}
\end{center}
\end{table}

\subsection{Results}

Table~\ref{tab:results} presents the main evaluation
results. For each scenario we report the number of PII
entity types exposed to the external model under the
Presidio baseline and under CAMP at $\tau = 2.0$, the
final CPE score at session end, and the turn at which
CAMP triggered retroactive pseudonymization.

\begin{table}[htbp]
\begin{center}
\caption{Evaluation Results at $\tau = 2.0$}
\label{tab:results}
\renewcommand{\arraystretch}{1.3}
\begin{tabular}{p{0.6cm} p{1.2cm} p{1.0cm} p{1.2cm}
p{1.5cm}}
\hline
\textbf{ID} & \textbf{Presidio Exp.} &
\textbf{CAMP Exp.} & \textbf{Final CPE} &
\textbf{Trigger} \\
\hline
S1 & 5 & 0 & 6.82 & Turn 4 \\
S2 & 6 & 0 & 9.25 & Turn 2 \\
S3 & 3 & 0 & 8.03 & Turn 2 \\
S4 & 5 & 0 & 7.48 & Turn 5 \\
\hline
\end{tabular}
\end{center}
\end{table}

\begin{figure}[htbp]
\centering
\includegraphics[width=\columnwidth]{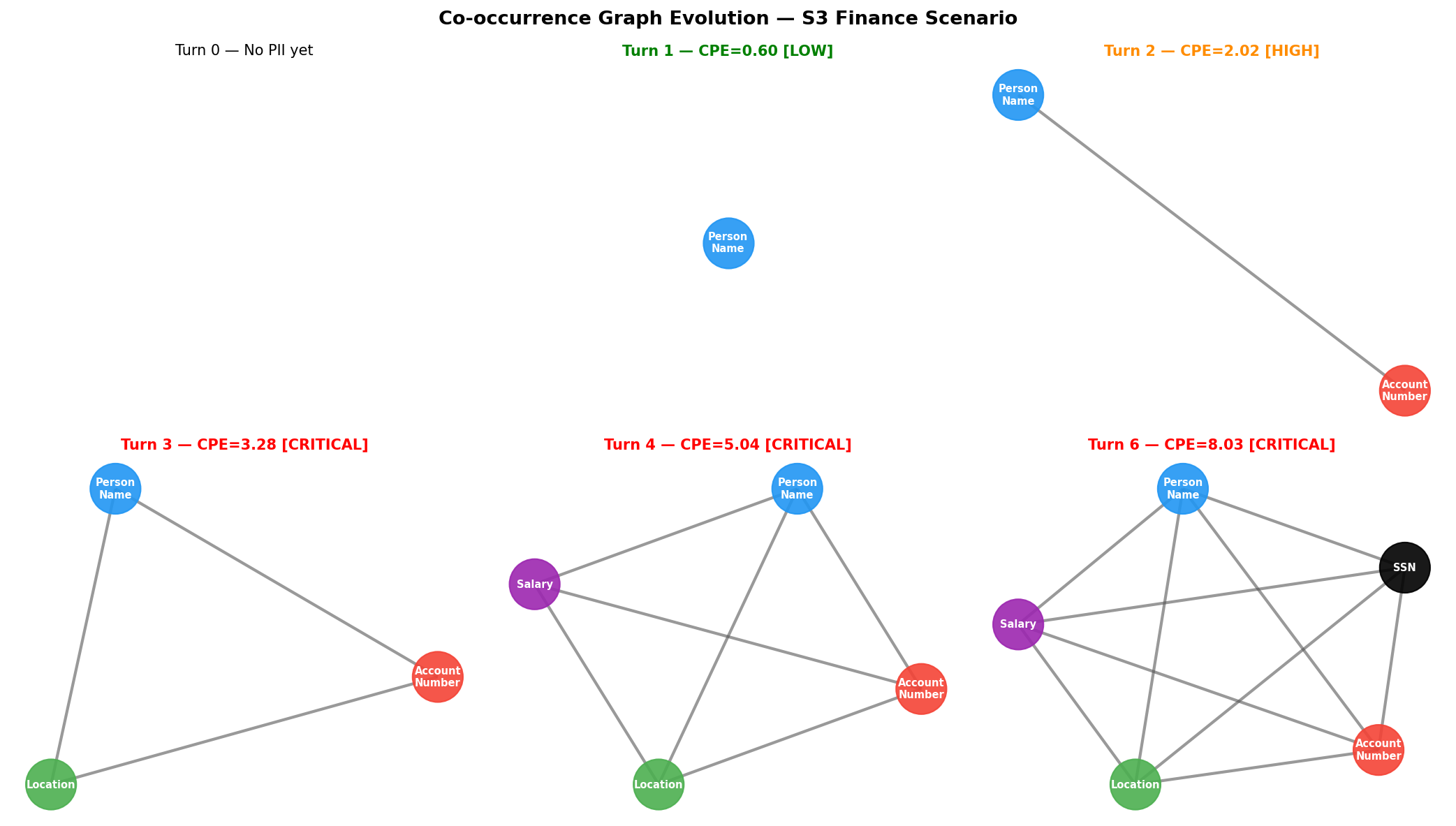}
\caption{Co-occurrence graph evolution across turns in
S3 Finance. Nodes represent PII entity types. Edges
form when two types co-occur in the same session. Graph
density drives the CPE amplifier.}
\label{fig:graph_evolution}
\end{figure}

Across all four scenarios CAMP reduced real PII entity
exposure to the external model to zero, while the
Presidio baseline forwarded between three and six
distinct entity types per session. The highest final
CPE score was observed in S2 at 9.25, reflecting rapid
accumulation of high-weight professional and demographic
identifiers including name, location, salary, age, and
ethnicity within seven turns. S3 reached the
intervention threshold earliest at turn~2, consistent
with the fast accumulation of financial identifiers.
S4 produced the most gradual accumulation pattern,
triggering at turn~5 across casual conversation turns
where no single message would raise concern under
per-turn inspection.

\subsection{Threshold Sensitivity}

Table~\ref{tab:threshold} reports the trigger turn
across all four scenarios under three threshold
settings. Lower thresholds cause earlier intervention,
providing stronger privacy guarantees at the cost of
pseudonymizing a greater portion of the conversation.
Higher thresholds allow more natural context to reach
the model before intervention, preserving utility at
the cost of later protection.

\begin{table}[htbp]
\begin{center}
\caption{Trigger Turn by Threshold Setting}
\label{tab:threshold}
\renewcommand{\arraystretch}{1.3}
\begin{tabular}{p{0.8cm} p{1.5cm} p{1.5cm} p{1.5cm}}
\hline
\textbf{ID} & $\boldsymbol{\tau = 1.5}$ &
$\boldsymbol{\tau = 2.0}$ &
$\boldsymbol{\tau = 2.5}$ \\
\hline
S1 & Turn 4 & Turn 4 & Turn 4 \\
S2 & Turn 2 & Turn 2 & Turn 2 \\
S3 & Turn 2 & Turn 2 & Turn 3 \\
S4 & Turn 3 & Turn 5 & Turn 5 \\
\hline
\end{tabular}
\end{center}
\end{table}

\begin{figure}[htbp]
\centering
\includegraphics[width=\columnwidth]{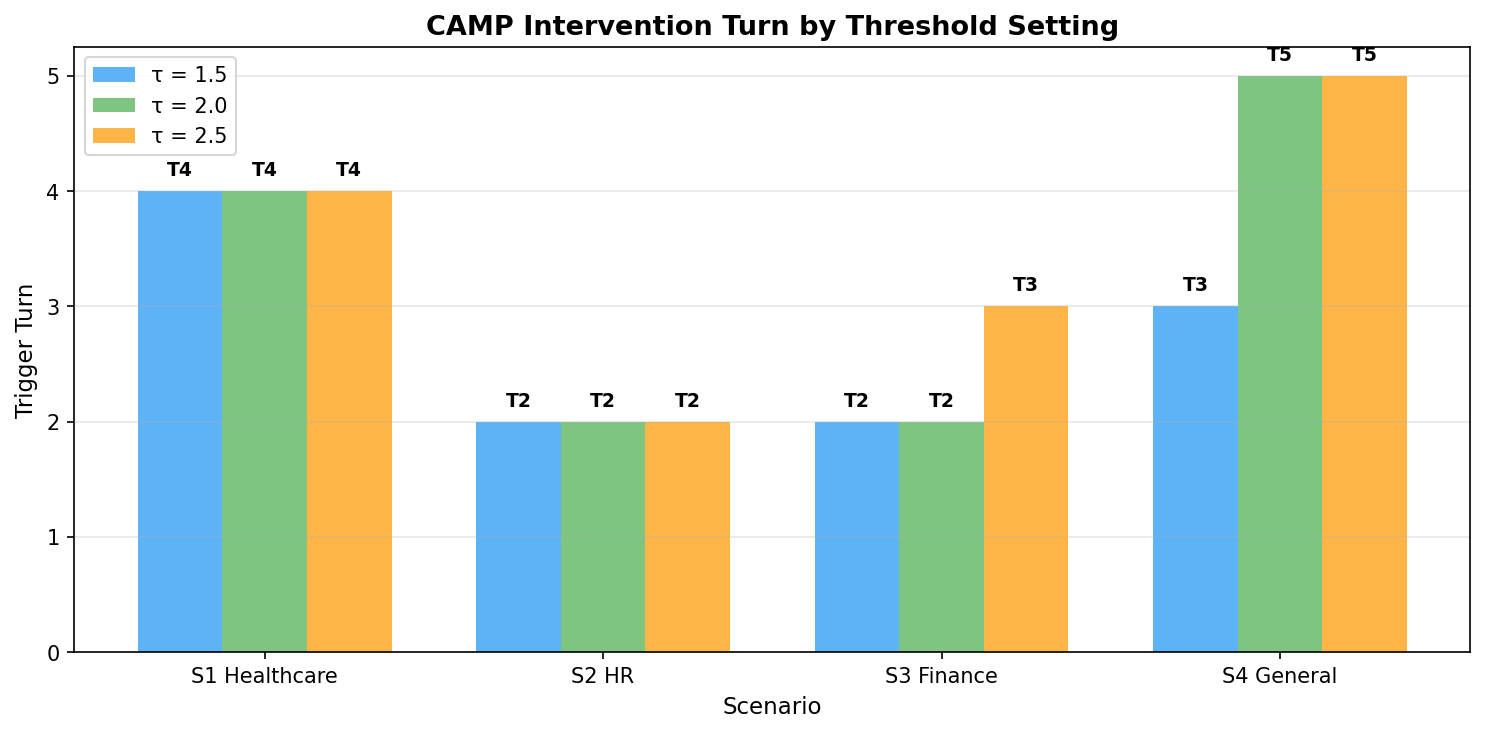}
\caption{CAMP intervention turn across all scenarios
under three threshold settings. Lower thresholds trigger
earlier intervention at the cost of pseudonymizing more
conversational context.}
\label{fig:threshold}
\end{figure}

The results indicate that $\tau = 2.0$ provides a
reasonable balance across all four domains. S1
Healthcare shows identical trigger turns across all
three thresholds at turn~4, suggesting the healthcare
disclosure pattern has a natural accumulation cliff at
that point regardless of threshold. S3 Finance is the
most threshold-sensitive scenario, shifting from turn~2
to turn~3 between $\tau = 2.0$ and $\tau = 2.5$. S4
General shows the widest variation, triggering at
turn~3 under $\tau = 1.5$ and turn~5 under both
$\tau = 2.0$ and $\tau = 2.5$, reflecting its slower
quasi-identifier accumulation pattern across casual
conversation.

\subsection{Utility Preservation}

In all four scenarios the LLM produced coherent,
contextually appropriate responses after
pseudonymization. The de-masking layer successfully
restored original identities in all responses, and no
synthetic substitutes were observed in the final output
shown to the user. This confirms that retroactive
pseudonymization does not degrade conversational utility
provided the pseudonyms are internally consistent across
turns, which the Faker-based pseudonymizer guarantees
by design.

\section{Discussion}

The experimental results demonstrate that per-turn
masking systems such as Presidio~\cite{presidio2021},
while effective at removing explicit identifiers within
individual messages, are structurally incapable of
addressing the privacy risk that emerges from gradual
PII accumulation across a conversation session. The
core issue is architectural: a stateless system that
processes each message independently cannot reason
about the combinatorial risk created by the intersection
of quasi-identifiers across turns. CAMP addresses this
by maintaining session state, modeling combination risk
explicitly through the co-occurrence graph, and
intervening retroactively when the accumulated risk
crosses a threshold.

The threshold parameter $\tau$ plays a central role in
governing the privacy-utility tradeoff. Our results
across $\tau \in \{1.5, 2.0, 2.5\}$ suggest that no
single value is universally optimal. Domains with
high-sensitivity data such as healthcare and finance
may warrant lower thresholds, accepting earlier
intervention to provide stronger guarantees. General
conversation assistants may benefit from higher
thresholds that allow more natural context to accumulate
before pseudonymization is applied. We recommend that
operators select $\tau$ based on the sensitivity profile
of their deployment domain rather than applying a single
fixed value across all use cases.

The combination amplifier parameter $\alpha$ controls
how aggressively the co-occurrence graph amplifies
combination risk. In our experiments we fixed
$\alpha = 0.3$, which produced reasonable sensitivity
across all four scenarios. Higher values of $\alpha$
would cause the CPE score to grow faster as the graph
becomes more connected, triggering earlier intervention.
A systematic study of the sensitivity of results to
$\alpha$ is an avenue for future work.

One important limitation of the current framework is
its reliance on explicit entity detection. CAMP
intervenes only on PII that Presidio successfully
identifies. Contextually embedded or implicit
disclosures, such as a user describing a workplace
without naming their employer, may not be detected
and therefore do not contribute to the CPE score.
This represents a fundamental limitation shared with
all NER-based approaches~\cite{lison2021anonymisation}
and motivates future exploration of LLM-based inference
as a complementary detection layer, provided the
inference itself is conducted locally to avoid the
circularity of sending PII to a model to detect PII.

A further consideration is the quality of synthetic
substitutes generated by the pseudonymizer. In our
implementation we use the Faker library which produces
plausible but generic substitutes. In domains where
the LLM response quality depends on the realism of
the pseudonyms, more sophisticated pseudonymization
strategies that preserve domain-specific properties
may be warranted, as suggested by recent work on
utility-preserving anonymization~\cite{rupta2025}.

\section{Future Directions}

Several avenues extend naturally from the foundation
established by CAMP.

\textbf{Adaptive threshold selection.} The current
framework requires the operator to configure $\tau$
manually. A natural extension would be to learn the
threshold adaptively from deployment feedback,
adjusting it based on observed re-identification risk
in a given domain without operator intervention.

\textbf{Inference-based implicit PII detection.} CAMP
currently detects only explicitly stated PII. A local
small language model trained to infer implicit
disclosures from conversational context could extend
detection coverage to quasi-identifiers that are never
directly stated but can be inferred from accumulated
context, extending the adversarial inference ideas
of~\cite{staab2024beyond} to the detection side.

\textbf{Multi-agent session tracking.} In multi-agent
architectures where a user interacts with an
orchestrator that delegates to specialized sub-agents,
PII accumulated in one agent context may be forwarded
to another~\cite{agentleak2026}. Extending CAMP to
track PII accumulation across agent boundaries rather
than within a single session is an important direction
for enterprise agentic deployments.

\textbf{Differential privacy integration.} The current
pseudonymization approach provides heuristic privacy
guarantees through consistent synthetic substitution.
Integrating formal differential privacy mechanisms at
the session level would provide stronger theoretical
guarantees about the relationship between the
pseudonymized context and the original identity.

\textbf{Evaluation on real conversational data.} Our
current evaluation relies on synthetic scenarios
designed to represent realistic disclosure patterns.
Evaluation on real anonymized conversational datasets
would provide stronger empirical validation of the CPE
metric and the intervention mechanism across a broader
range of natural disclosure patterns.

\textbf{Regulatory alignment.} GDPR Article~25 mandates
data minimization and privacy by design as core
principles for systems that process personal data.
Formalizing the relationship between the CPE threshold
and regulatory data minimization requirements would
provide a principled basis for operators to configure
CAMP in compliance with applicable data protection law.

\section{Conclusion}

This research introduced CAMP, a cross-turn privacy
protection framework for multi-turn LLM conversations
that addresses a class of privacy vulnerabilities not
covered by existing per-turn masking systems such as
Presidio~\cite{presidio2021} and
PAPILLON~\cite{siyan2024papillon}. We identified and
formalized Cumulative PII Exposure as a session-level
privacy risk metric, demonstrating that the gradual
accumulation of individually innocuous PII fragments
across conversation turns creates re-identifiable
profiles that stateless masking systems cannot detect
or prevent.

The key insight driving CAMP is that privacy risk in
multi-turn conversations is combinatorial rather than
additive. A location entity, a salary disclosure, and
an employer name are each weak identifiers in isolation.
Their co-occurrence within a session context window
forwarded to an external LLM creates a combination that
is substantially more dangerous than the sum of its
parts. The co-occurrence graph and CPE score provide a
principled mechanism for quantifying this combination
risk and triggering intervention at the appropriate
moment, extending prior work on inference-based privacy
risks~\cite{staab2024beyond,nakka2024piiscope} to the
multi-turn conversational setting.

CAMP addresses this risk through retroactive
pseudonymization of the full conversation history when
the CPE score crosses a configurable threshold,
replacing real PII values with consistent synthetic
substitutes before forwarding context to the external
model. A de-masking layer restores original identities
in the LLM response before it is returned to the user,
completing a full privacy-preserving round trip in
which the external model processes only synthetic
identities throughout the session.

Evaluation across four synthetic multi-turn scenarios
spanning healthcare, hiring, finance, and general
conversation demonstrated that CAMP reduced real PII
entity exposure to the external model to zero across
all scenarios, while the Presidio per-turn baseline
forwarded between three and six distinct entity types
per session. Conversational utility was fully preserved
in all cases, confirming that the privacy-utility
tradeoff inherent in aggressive per-turn
masking~\cite{intact2024,adaptivepii2025} can be
avoided through session-aware intervention.

We hope that the formalization of CPE and the CAMP
framework provide a useful foundation for future
research on privacy in agentic LLM
systems~\cite{agentleak2026,privacyguard2025}, and
that the session-level perspective introduced here
encourages broader consideration of cross-turn privacy
risk in the design of conversational AI deployments.

\onecolumn
\bibliographystyle{IEEEtran}
\bibliography{camp}

\end{document}